%% file: main.tex
\documentclass[dvipsnames, 12pt]{article}
\usepackage[margin=25mm]{geometry}
\usepackage{amsmath}\allowdisplaybreaks
\usepackage{amsfonts}
\usepackage{amssymb}
\usepackage{verbatim}
\usepackage{setspace}
\usepackage{rotating}
\usepackage{booktabs}
\usepackage{longtable}
\usepackage{tabularx}
\usepackage[x11names]{xcolor}
\usepackage{subcaption}
    \usepackage[flushleft]{threeparttable}
\usepackage{colortbl}
\usepackage{svg}
\usepackage[section]{placeins}
\usepackage{float}

\usepackage{etoolbox}\AtBeginEnvironment{thebibliography}{\linespread{1}\selectfont}
\newcommand{\zerodisplayskips}{%
  \setlength{\abovedisplayskip}{-5pt}%
  \setlength{\belowdisplayskip}{5pt}%
  \setlength{\abovedisplayshortskip}{-5pt}%
  \setlength{\belowdisplayshortskip}{5pt}}
\appto{\normalsize}{\zerodisplayskips}
\appto{\small}{\zerodisplayskips}
\appto{\footnotesize}{\zerodisplayskips}
\usepackage{tikz}
\usetikzlibrary{fit, arrows.meta, positioning}
\usetikzlibrary{shapes.geometric, arrows, decorations.pathreplacing,calc}

\tikzstyle{circleblock} = [circle, draw, minimum size=2.5cm, text centered]
\tikzstyle{rectangleblock} = [rectangle, draw, minimum width=4.5cm, minimum height=1.5cm, text centered]
\tikzstyle{smallrectangle} = [rectangle, draw, minimum width=3cm, minimum height=0.8cm, text centered]
\tikzstyle{arrow} = [thick,->,>=stealth]

\usepackage{amsthm}
\usepackage{bm}
\usepackage{dsfont}
\usepackage{lscape}
\usepackage{geometry}
\usepackage{float}
\usepackage{verbatim}
\usepackage[toc,page,header]{appendix}
\usepackage{minitoc}

\usepackage{dsfont}
\usepackage{tablefootnote}
\usepackage{multirow}
\usepackage{mathtools}
\usepackage{pdflscape}

\usepackage{algorithm}
\usepackage{algpseudocode}

\algdef{SE}[SUBALG]{Indent}{EndIndent}{}{\algorithmicend\ }%
\algtext*{Indent}
\algtext*{EndIndent}

\newcommand\Algphase[1]{%
\vspace*{-.7\baselineskip}\Statex\hspace*{\dimexpr-\algorithmicindent-2pt\relax}\rule{\textwidth}{0.4pt}%
\Statex#1
\vspace*{-.7\baselineskip}\Statex\hspace*{\dimexpr-\algorithmicindent-2pt\relax}\rule{\textwidth}{0.4pt}%
}

\usepackage{hyperref}
\hypersetup{
    colorlinks=true,
    linkcolor=cyan,
    filecolor=cyan,      
    urlcolor=cyan,
    citecolor=cyan,
    breaklinks=true}

\usepackage[square,numbers]{natbib}
\usepackage{authblk}
\usepackage{caption}
\usepackage{framed}
 
\usepackage{listings}

\usepackage{tcolorbox}
\usepackage{varwidth}

\begin{document}
\doparttoc 

\title{\textbf{A Bayesian approach to aggregated chemical exposure assessment}}

\author[1,2,$\ast$]{Sophie Van Den Neucker}
\author[1]{Alexander Grigoriev}
\author[2]{Heidi Demaegdt}
\author[2]{Jan Mast}
\author[2]{Karlien Cheyns}
\author[3]{Sofie De Broe}
\author[4]{Roberto Cerina}
\affil[1]{Data Analytics and Digitalisation, Maastricht University}
\affil[2]{Trace Elements and Nanomaterials, Sciensano}
\affil[3]{Data Governance, Sciensano}
\affil[4]{Institute for Logic, Language and Computation, University of Amsterdam}
\affil[*]{Corresponding author: \href{email:sophie.vandenneucker@maastrichtuniversity.nl}{sophie.vandenneucker@maastrichtuniversity.nl.}}

\date{}
\maketitle
\thispagestyle{empty}

\begin{abstract}
\noindent
Human exposure to chemicals commonly arises from multiple sources, yet traditional assessments often treat these sources in isolation, overlooking their combined impact. We introduce a Bayesian framework for aggregated chemical exposure assessment that explicitly accounts for these intertwined pathways. By integrating diverse datasets --- such as consumption surveys, demographics, chemical measurements, and market presence --- our approach addresses typical data challenges, including missing values, limited sample sizes, and inconsistencies, while incorporating relevant prior knowledge. Through a simulation-based strategy that reflects the full spectrum of individual exposure scenarios, we derive robust, population-level estimates of aggregated exposure. We demonstrate the value of this method using titanium dioxide, a chemical found in foods, dietary supplements, medicines, and personal care products. By capturing the complexity of real-world exposures, this comprehensive Bayesian approach provides decision-makers with more reliable probabilistic estimates to inform public health policies.
\end{abstract}

\newpage
\pagenumbering{arabic} 

\newpage
\section{Introduction} \label{sec1}
In our daily lives, we are exposed to hundreds and thousands of different chemicals from a wide range of sources. They are in the food that we eat, the air that we breathe, and the products that we use. While many of these chemicals are safe for everyday use and play an essential role in various industries, some can be harmful to our health --- resulting in acute (e.g., respiratory irritation) or chronic effects (e.g., cancer). To assess the potential risk that a chemical poses to public health, it is not only essential to determine whether it is harmful and at what level, but also to understand the extent we are actually exposed to it as a population. This is exemplified by the fact that some harmful chemicals can be considered as not posing a risk if the exposure to them is very low.\\

Traditionally, chemical exposure assessments performed for regulatory purposes focus on a single chemical (e.g., titanium dioxide) from one particular source (e.g., food). While this approach is methodologically more straightforward, it disregards the fact that the various chemical exposures from our environment do not act in isolation, but rather impact our health in an interconnected manner. 
This perspective is derived from the \emph{exposome} paradigm, which aims to measure the totality of a person's exposures throughout their life and the relationship of these exposures to their health \cite{wild_2005}.
Not only are we exposed to a variety of chemicals, but we are also often exposed to a particular chemical from multiple sources via different pathways (e.g., ingestion, inhalation, dermal contact). An example is titanium dioxide (TiO$_2$) --- a chemical found in food, dietary supplements, medicines, personal care products, and other applications. The sum of these exposures is generally referred to as \emph{aggregated exposure}, and understanding it is essential for assessing any related health risks. Since this is often not structurally taken into account, a chemical may be considered safe when assessed within the regulatory framework of each of its sources separately, even though the exposure across these sources exceeds acceptable limits --- thus potentially underestimating the associated risk \cite{hermans_2024}. \\

Assessment of aggregated exposure is particularly of interest when a chemical is found in a variety of products available on the market to which the general population is regularly exposed. Many existing approaches to this purpose focus on a single industry (e.g., cosmetics) and only consider worst-case scenarios, resulting in an overestimation of the associated risk (\citealp[e.g.,][]{sccs_2023}). However, recent methodological advancements have increasingly been incorporating more probabilistic approaches that consider the full spectrum of possible individual exposure scenarios (e.g., PACEM; \citealp{delmaar_2023}). These commonly rely on (non-)parametric methods based on frequentist principles (e.g., bootstrap) used to randomly sample individual quantities in a Monte Carlo simulation. While these approaches have many advantages, they can lack the flexibility that is required to accurately represent the data-generating process underlying such a complex reality.\\

In this paper, a conceptual framework is proposed for the aggregated exposure assessment of chemicals using Bayesian inference. The aim is to hereby demonstrate how an inherently probabilistic alternative to existing approaches is more flexible and more easily adaptable to a wide range of contexts. This framework is illustrated using TiO$_2$ as it is a chemical found in a variety of products available to the general public across different industries. Although long considered safe for human health, there has been some disagreement on the subject in recent years. In 2022, it was banned as a food additive in the European Union due to concerns that it might be toxic to humans when ingested \cite{efsa_2021}. This decision has not yet been echoed in other parts of the world or industries; however, discussions are ongoing \cite{ema_2021}. It is therefore of interest to consider the aggregated exposure across its different sources to obtain a more comprehensive understanding of the potential health risks.

\section{Estimating long-term aggregated exposure} \label{sec2}

\begin{figure}[H]
    \centering
    \includegraphics[scale = 0.43]{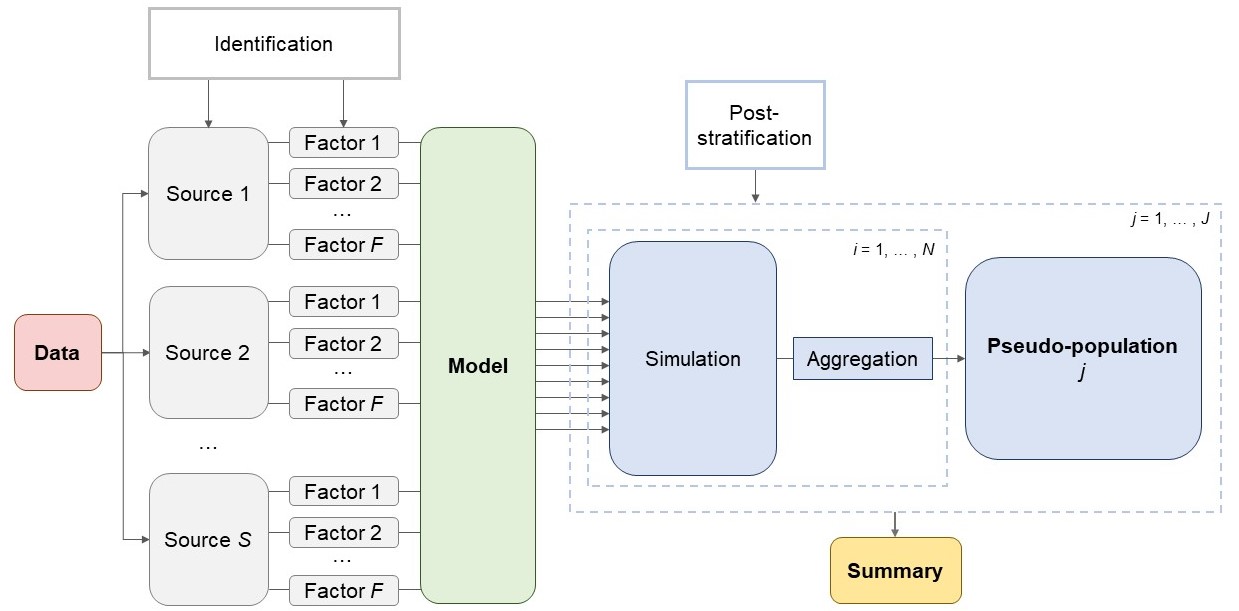}
    \caption{Schematic representation of the proposed framework for estimating long-term aggregated chemical exposure.}
    \label{fig1}
\end{figure}

The proposed approach for estimating long-term aggregated exposure to a particular chemical in a Bayesian framework involves a series of consecutive steps to derive accurate and robust estimates at the population level, as shown schematically in Figure~\ref{fig1}. In what follows, a general description of these steps is given as a basis for application to a broad range of contexts. The focus was hereby narrowed down to long-term exposure of products available on the market, representing the average daily amount of a chemical that an individual is exposed to from using or consuming these products over a long period of time (i.e. at least one year).

\emph{First}, the \textbf{identification} of all the \emph{sources} wherein a chemical of interest can be found is essential to assess the aggregated exposure to it. These sources can be determined at varying hierarchical levels depending on the context. For example, food can be considered as an overarching source, although it includes a range of subcategories. A distinction is herein made with exposure \emph{factors}, such as the \emph{frequency} and \emph{amount} of exposure to a particular source, the chemical \emph{concentration} in the products belonging to this source, their \emph{market presence}, the \emph{body weight} of the exposed individual, and any factors related specifically to the pathway of exposure (e.g., ingestion rate). These factors contribute in one way or another to the extent to which an individual is exposed to the chemical in question.\\

\emph{Second}, in the common absence of one \textbf{data} source encompassing information on each of these factors per exposure source at the individual level, distinct data sources need to be retrieved. Let $y^{s_f}$ be the outcome of interest for factor $f$ of exposure source $s$. The nature of the information that is required, its level of granularity, its subject --- individual or product --- and where it can be found depend on the target. 

Data on the \emph{amount and frequency of exposure} are often derived from surveys on individuals’ consumption or usage of a particular source (e.g., food). Participants can be asked via a questionnaire to estimate how often and how much they are exposed to a certain category of products belonging to that source (e.g., confectionery). Alternatively, they can be asked via an interview to recall this information more precisely for the last 24 hours, which is then repeated over the course of multiple days. These surveys generally contain information on the individual characteristics of their participants, such as age, gender, place of residence, and body weight, which is essential to derive exposure estimates for demographic groups of interest. 

Data on the \emph{concentration} of the chemical of interest in these products is usually derived from analytical measurements obtained through various methods (e.g., spectroscopy) executed in a controlled laboratory environment. They represent the amount of a certain chemical compound (e.g., TiO$_2$), or one of its elements (e.g., Ti), that is present in a product sample. Use levels reported by industry of the chemical concentrations in their products can function as a surrogate or complement in this regard. 

Data on the \emph{market presence} of a particular chemical, i.e. the proportion of products in a given category that contain the chemical of interest (e.g., chewing gum), can, in principle, be derived from product composition databases (e.g., Mintel Global New Products Database). These generally describe the main characteristics of individual products available for purchase to the general public, such as their ingredient list. 

However, distinct data sources tend to describe individuals and products differently. For example, the products reported in a consumption or usage survey may be classified according to one system (e.g., FoodEx2) and those described in a product composition database according to another.  It is therefore essential to link them in a comprehensive manner prior to modeling, as failing to do so would hinder the combination of their respective parameter estimates during post-processing.\\

\emph{Third}, an appropriate \textbf{model} must be developed for each exposure factor per exposure source in accordance with the data that was collected. Let $p(y^{s_f}|\theta^{s_f})$ be the sampling distribution of $y^{s_f}$ given parameters $\theta^{s_f}$. The general functional form of $p(y^{s_f}|\theta^{s_f})$ is often characteristic of typical continuous distributions for the amount and the concentration (e.g., Gamma, Lognormal) and discrete ones for the frequency and the market presence (e.g., Binomial). However, specific model definitions, including transformations (e.g., Box-Cox), finite mixtures, and the inclusion of longitudinal and spatial effects, are context dependent. By incorporating relevant predictors such as age group, gender, place of residence, and socioeconomic status, estimation in demographic groups that are of interest from a public health perspective is enabled.  

Let $p(\theta^{s_f})$ be the joint prior probability distribution of $\theta^{s_f}$ incorporating any information about these parameters independently of the data --- including context-specific expert knowledge. 
The data collected for an exposure assessment is often grouped in various ways --- observations within individuals (e.g., repeated 24-hour interviews in a survey), individuals within geographical areas, products within categories (e.g., chewing gum $>$ confectionery $>$ food). Including this structure in the model allows to account for any variability in the outcome of interest and derive estimation at different levels. However, it is often the case that there is more data available in some groups than in others. This information can be leveraged by treating some of the parameters $\phi^{s_f}$ that define $p(\theta^{s_f})$ as unknown and assigning them their own prior $p(\phi^{s_f})$. These hierarchical priors $p(\theta^{s_f}|\phi^{s_f})$ introduce dependence between groups so that they can implicitly borrow strength from each other \cite{gelman_2006}. This is a form of regularization, which reduces overfitting and improves generalizability.  Furthermore, when dealing with a large amount of data across multiple datasets, it is inevitable to come across missing values, be it in the outcomes or predictors. Body weight is one example of a variable often plagued by missingness owing to various reasons, such as survey participants’ reluctance to disclose personal information. However, it is an essential factor in chemical exposure assessments, since the outcome of interest is generally expressed in mg of the chemical substance per kg of body weight per day (`mg/kg/day’). Adjusting for it ensures comparability between different demographic groups based on age and gender. Instead of excluding these missing values entirely or imputing them deterministically, they can be imputed probabilistically. This involves treating them as parameters and assigning them a prior informed by the observed values $y_{\mbox{\small{obs}}}^{s_f}$. The missing values $y_{\mbox{\small{mis}}}^{s_f}$ are then imputed iteratively by their posterior estimates derived from $p(y_{\mbox{\small{mis}}}^{s_f}|y_{\mbox{\small{obs}}}^{s_f}, \theta^{s_f})$. Note that this inherently assumes that these values are missing (completely) at random \cite{little_2021} and are therefore exchangeable, conditional on the predictors included in the model. Although conceptually different, this approach can also be applied to account for uncertainty inherent to the data, such as analytical measurements used to determine the chemical concentration in a product. This information (e.g., reported standard errors) can be incorporated into the model through the prior in a similar manner and will thus be propagated in the posterior \cite{richardson_1993}. The available data often does not include information on more than one factor per exposure source for the same subject. However, when it does, their correlation can be estimated in a multivariate model. This can be achieved by including a joint prior on subject-specific effects particular to each factor, like the amount and frequency of food consumed. Suppose $\alpha$ and $\beta$ are these effects derived from the respective likelihoods of factor $A$ and factor $B$ such that, assuming a normal density, the joint prior can be defined as $\alpha, \beta \sim \mbox{Normal}(0, \Sigma_{\alpha, \beta})$.

Let $p(\theta^{s_f}|y^{s_f})$ be the joint posterior probability distribution from which samples of the parameters $\theta$ are drawn iteratively using a suitable Markov Chain Monte Carlo (MCMC) algorithm. One example is the No-U-Turn Sampler (NUTS), an adaptive variant of the Hamiltonian Monte Carlo algorithm \cite{nuts}. It distinguishes itself from other MCMC algorithms, such as Metropolis and Gibbs, in its ability to handle highly complex models in a more computationally efficient manner. This is particularly pertinent when considering multiple factors across various sources, often requiring thousands of parameters to be estimated jointly.\\

\emph{Fourth}, a \textbf{pseudo-population} reflecting the full spectrum of possible individual exposure scenarios is generated in a simulation-based approach to compensate for the fact that real data on aggregated exposure in the population often does not exist.  Let $\phi^{s_f^{(j)}}$ and $\theta^{s_f^{(j)}}$ be the values of the corresponding parameters drawn from the joint posterior distributions $p(\phi^{s_f}|y^{s_f})$ and $p(\theta^{s_f}|\phi^{s_f}, y^{s_f})$, respectively, at iteration $j$ of the sampling phase of the MCMC algorithm (excluding samples from the warmup phase). The long-term exposure of a pseudo-individual $i$ in population $j$ to source $s$ is determined by computing and combining values for the corresponding long-term average $\tilde{\mu}^{s_f^{(j)}}$ of each factor from the models defined previously given $\phi^{(j)}$ and $\theta^{(j)}$. Depending on the purpose, this involves simulating values $\tilde{y}^{(j)}$ and $\tilde{\theta}^{(j)}$ from $p(y^{s_f}|\theta^{s_f^{(j)}})$ and $p(\theta|\phi^{s_f^{(j)}})$, respectively,  to generate the quantities of interest. The long-term aggregated exposure of pseudo-individual $i$ in population $j$ across $S$ sources is then obtained as follows:

\begin{equation}
  \tilde{\mu}_i^{(j)} = \sum_{s = 1} \frac{\prod_{f = 1} \tilde{\mu}^{{s_f^{(j)}}}_{i}}{\tilde{y}_i^{w(j)}}, \hspace{0.5cm} \mbox{for } j = 1 \mbox{, \ldots, } J; f = 1 \mbox{, \ldots, } F;  s = 1 \mbox{, \ldots, } S  \label{eq1}
\end{equation}
where $\tilde{y}_i^{w(j)}$ is the body weight of pseudo-individual $i$.
When repeated for a large number of pseudo-individuals, the inherent randomness of the simulation ensures, in principle, that the pseudo-population incorporates the variability found in real-world individual exposure scenarios. Suppose, for example, that two similar individuals consume chewing gum. One does so frequently, but from brands not using the chemical of interest in their products, the other infrequently, but from brands that do use it. As such, considering only one source (i.e. food) and two factors (i.e. frequency and market presence) for simplicity, the exposure of the latter will be higher than that of the former. Through the above simulation approach, these two scenarios will be represented in the pseudo-population according to their respective probability of occurring as determined by the modeling assumptions and available data. In order to draw meaningful conclusions about the true population of interest from the generated pseudo-population, the latter must be representative of the former regarding the outcome of interest. However, the data that is used to fit the models described above are often derived from samples that exhibit some degree of non-representativeness. For example, it is common in surveys to disproportionally sample more individuals from a particular subpopulation to increase precision about the corresponding parameter estimates of interest. To adjust for any observed non-representativeness in the data and thus avoid potential bias, a post-stratification scheme can be incorporated when generating the pseudo-population. This requires information on the joint distribution of several key demographic indicators, such as age group, gender, and place of residence, which can be obtained directly from census data or estimated from other sources \cite[e.g.,][]{cerina_2021}. For pseudo-population $j$, it implies simulating the long-term aggregated exposure for a number of individuals per demographic group or stratum (e.g., `12-17 years' --- `male' --- `Brussels') proportionally to the corresponding number in the true population. As such, the overall generated pseudo-population is, in principle, representative of its true counterpart. It should be noted that post-stratifying is only possible if information on these demographic indicators is available in the data and adjusted for during the modeling of the different exposure factors to simulate corresponding quantities of interest. Furthermore, when combining post-stratification with hierarchical priors, this approach is commonly referred to as multilevel regression and post-stratification \cite{downes_2018}. \\

\emph{Fifth}, from the resulting empirical distribution of the long-term aggregated exposure values across all pseudo-individuals, \textbf{summary} statistics of interest like the median and 2.5th and 97.5th percentiles can be computed. When repeating this approach for every pseudo-population $j$, a sample of size $J$ from the posterior of these parameters is obtained, quantifying uncertainty about them given the data and the designated model assumptions.

\section{Titanium dioxide as a case study} \label{sec3}

To illustrate the approach described above, a long-term aggregated exposure assessment of TiO$_2$ from food, dietary supplements, medicines, and personal care products, i.e. toothpaste and lip balm, was performed for the Belgian population in the period of 2014 -- 2024.

\subsection{Data sources} \label{sec3.1}

A description of the different data sources used for the long-term aggregated exposure assessment of TiO$_2$ is given in Table~\ref{tab1}.

\begin{sidewaystable}[!htp]\centering
\caption{A summary of the data sources used to estimate the long-term aggregated exposure to TiO$_2$}
\label{tab1}
\footnotesize
\begin{tabular}{p{0.5in}p{1.8in}p{1.6in}p{0.9in}p{0.5in}p{0.8in}p{1.4in}}
&&&&&&\\
\toprule
\textbf{Name} &\textbf{Description} &\textbf{Source} &\textbf{Time period} &\textbf{Region} &\textbf{Sample size} &\textbf{Unit of  observation}  \\
\midrule
CON &Concentrations of TiO$_2$ in consumer products & \cite{vandenneucker_2025} &2000-2024 &World &674 &Product \\
DEM &Population numbers by demographic group &Statbel (the Belgian statistical office) &2015 &Belgium & 3,477 &Province  \\
FAR &Characteristics of lip balms available on the market (e.g., ingredient list) &Farmaline (online store of pharmaceutical and personal care products; data extracted automatically for the present paper)  &2024 &Belgium &211 &Product  \\
FCS & The type and amount of food consumed by survey participants on two non-consecutive interview days & Food Consumption Survey  \cite{bel_2016} &2014-2015 &Belgium &3,200 &Individual \\
HIS & The type and frequency of medicines used by survey participants &Health Interview Survey  \cite{demarest_2018} &2018 &Belgium &10,700 &Individual  \\
MED &Characteristics of authorized medicines (e.g., active substance) &Federal Agency for Medicines and Health Products &2024 &Belgium &11,180 &Product  \\
OFF & Characteristics of food products available on the market (e.g., ingredient list) &Open Food Facts (crowdsourced database) &2024 &World & 2,948,707 &Product  \\
PCP &Survey on the consumption of personal care products  & \cite{ficheux_2017} &2013-2016 &France & &Product  \\
SUP & Characteristics of toothpastes available on the market (e.g., ingredient list)  &Belgian supermarkets (data extracted manually for the present paper) &2023-2024 &Belgium &88 &Product \\
\bottomrule

\end{tabular}
\end{sidewaystable}

\subsection{Modeling approach} \label{sec3.2}

Since every factor plays an essential role in the estimation of long-term exposure, a suitable model was developed in accordance with the available data and including relevant predictors as appropriate. The corresponding mathematical model formulation and underlying assumptions are described below by exposure source.  A more detailed description is given in Appendix A. The general notational convention that was used follows \cite{gelman_hill_2006}, with particular definitions shown in Table~\ref{tab2}. Owing to the inherent complexity of estimating aggregated exposure, a distinction was made between the role of superscripts and subscripts to simplify notation, with the former referring to labels and the latter to indices. For example, $y^{F_a}_{it}$ represents the amount ($a$) of food ($F$) consumed by individual $i$ on day $t$. 

\begin{table}[H]
    \centering
    \caption{The definition of the labels and indices used in the mathematical model formulations of the different exposure factors in the estimation of the long-term aggregated exposure to TiO$_2$.}
    \label{tab2}
    \begin{tabular}{llllll}
    \toprule
    \textbf{Label} &&&& \textbf{Index}\\
    \midrule 
       \textit{F}  &  food &&& \textit{g} & group/category\\
       \textit{M}  &  medicine &&& \textit{i}  &  individual\\
       \textit{P}  &  personal care &&& \textit{j}  &  posterior sample\\
       \textit{S}  &  dietary supplements &&& \textit{k}  &  predictor\\
       \textit{a}  &  amount &&& \textit{l}  &  level of predictor\\
       \textit{c}  &  concentration &&& \textit{p}  &  product\\
       \textit{f}  &  frequency &&& \textit{t}  &  day\\
       \textit{m}  &  market presence &&&  &  \\
       \textit{w}  &  weight &&&  &  \\
    \bottomrule
    \end{tabular}
\end{table}

\subsubsection{Food} 

The exposure \emph{frequency} of food containing TiO$_2$ was modeled using the FCS data (Food Consumption Survey data, see Table~\ref{tab1}) as follows

\begin{alignat}{2}
y^{F_f}_{git} &\sim \mbox{Bernoulli}(\pi^{F_f}_{gi})  &{} \mbox{ for }{}& i = 1, \dots, N^f \mbox{; } t \in \{1, 2\}  \label{eq2} \\
\mbox{logit}(\pi^{F_f}_{gi}) &= \eta^0_g + \sum_{k = 1}^K \eta^k_{gl_k[i]}  + \upsilon_{gi} &{} \label{eq3}
\end{alignat}
where $y^F_{f_{git}}$ in (\ref{eq2}) is a binary indicator of whether individual $i$ consumed one or more food products of a certain category $g$ on interview day $t$ (0: no; 1: yes); $\eta^0_g$ is the effect of food category $g$; $\eta^k_{gl_k}$ is the effect of the $l$th level of predictor $k$ varying by category; and $\upsilon_{gi}$ is an individual effect taking into account the repeated nature of the data over time. \\

The exposure \emph{amount} of food containing TiO$_2$ was coincidentally modeled using the FCS data as follows

\begin{alignat}{2}
\frac{({y^{F_a}_{git} / y^{F_a}_{w_i})}^{\lambda^F} - 1} {\lambda^F} &\sim \mbox{Normal} (\mu^{F_{a/w}}_{gi}, \sigma^{F_{a/w}}) &{}  \mbox{ for }{}& i = 1, \dots, N^a \mbox{; } t \in \{1, 2\} \label{eq4}\\
\mu^{F_{a/w}}_{gi} &= \gamma^0_g + \sum_{k = 1}^K \gamma^k_{gl_k[i]} + \nu_{gi} \label{eq5} 
\end{alignat}
where $y^F_{a_{git}}$ in (\ref{eq4}) is the total daily amount of it consumed (in g) if $y^F_{f_{git}} = 1$, such that this model only includes individuals with non-zero amounts; $\gamma^0_g$, $\gamma^k_{gl_k}$, and $\nu_{gi}$ in (\ref{eq5}) have a similar interpretation to $\eta^0_g$, $\eta^k_{gl_k}$, and $\upsilon_{gi}$ in (\ref{eq3}), respectively. An adjustment for the body weight  $y^{F_w}_i$ of individual $i$ (in kg) was included in (\ref{eq4}), with any missing values iteratively imputed according to 
\begin{alignat}{2}
\log(y^{F_w}_{i}) &\sim \mbox{Normal} (\mu^{F_w}_{i}, \sigma_w^{F_w}) \label{eq6}\\
\mu^{F_w}_{i} &= \beta_0 + \sum_{k = 1}^{K} \beta^{k.(k+1)}_{l_k[i], l_{k+1}[i]} \label{eq7} 
\end{alignat}
where $\beta^{k.(k+1)}$ is the effect of the interaction between predictors $k$ and $k+1$. The adjusted amounts were further transformed according to Box-Cox \cite{box_cox_1964} to support the assumption of a Normal likelihood, with $\lambda^F$ in (\ref{eq4}) estimated jointly with the other parameters of the model.

To explicitly account for any correlation between the probability of consuming a food from a certain category and the amount of it consumed, a joint hierarchical prior was defined for the individual effects $\upsilon_{gi}, \nu_{gi} \sim \mbox{Normal}(0, \Sigma_{\upsilon, \nu})$, with the variance and covariance between elements varying by food category following \cite{paulo_2005}. \\

The \emph{concentration} of TiO$_2$ in food was modeled as previously described (Van Den Neucker et al., 2025) using the CON data (Concentrations data, see Table~\ref{tab1}). In brief, it was assumed that the concentration in product $p$ of category $g$, namely $y^{F_c}_{gp}$, follows a Gamma distribution with a median $\tilde{\mu}^{F_c}_{gp}$ varying between three levels of food categories. Hierarchical priors were used to leverage information where there was data sparsity.\\ 

The \emph{market presence} was modeled using the OFF data (Open Food Facts data, see Table~\ref{tab1}) as $y^{F_m}_{p} \sim \mbox{Bernoulli}(\pi^{F_m}_{p})$, where $y^F_{m_p}$ is a binary indicator of whether product $p$ contains TiO$_2$ as an ingredient (0: no; 1: yes) and $\pi^{F_m}_{p}$ is the corresponding probability, which varies according to the food category to which it belongs to.

\subsubsection{Dietary supplements} 

Since the FCS data also contains information on the consumption of dietary supplements, the related exposure \emph{frequency} was modeled similarly to (\ref{eq3}), but adapted to the case of $G=1$. Accordingly, the exposure \emph{amount} of dietary supplements was modeled as follows

\begin{align}
\log\left (\frac{ \sum_t  y^{S_a}_{it} / T_i}{y^{S_w}_{i}} \right ) &\sim \mbox{Normal} (\mu^{S_{a/w}}_{z[i]}, \sigma^{S_{a/w}}_{z[i]}) &{} \mbox{ for }{}& i = 1, \dots, N^{F_a} \mbox{; } t \in \{1, 2\} \label{eq8} \\
\mu^{S_{a/w}}_{z[i]} &= \rho^0_{z[i]} + \sum_{k = 1}^K \rho^k_{l_k[i]} \label{eq9}\\
z_i &\sim \mbox{Categorical}(\vartheta_{z[i]}) &\mbox{for }& z = 1,.., 3 \mbox{ and } \sum^3_{z = 1} \vartheta_z = 1 \label{eq10}
\end{align}
where $y^{S_a}_{it}$ in (\ref{eq8}) is the total daily amount of dietary supplements consumed by individual $i$ on interview day $t$ (in g). It was averaged across the days on which there was a consumption event, i.e. $y^{S_f}_{it}$ = 1, given a total number of days $T_i$,  and then adjusted for the individual body weight $y^{S_w}_{i}$ (kg), which was modeled similarly to (\ref{eq7}).
Although the data used to estimate the exposure amount of dietary supplements is derived from the same source as for food, a different approach was required. Indeed, dietary supplements are generally consumed in integer amounts (e.g. 1 g), resulting in a multimodal empirical distribution even after adjusting for body weight and other observed variables. Thus, a finite mixture was used to account for the remaining variability in the outcome of interest. The latent categorical variable and corresponding proportion indicating which component characterizes individual $i$ are represented by $z_i$ and $\vartheta_{k_i}$ in (\ref{eq10}), respectively. In particular, the intercept $\rho^0_{z[i]}$ in (\ref{eq9}) was assumed to vary between the mixture components, while the effects of the included predictors, i.e. $\rho^k_{l_k[i]}$ in (\ref{eq9}), were not.  \\

The \emph{concentration} of TiO$_2$ was not explicitly modeled because of challenges arising from a small sample size and unobserved variability. Instead, the empirical distribution was sampled with replacement iteratively during post-processing.\\

The \emph{market presence} was modeled using the OFF data in a similar manner to the one described for the food, with $\pi^{S_m}_p$ representing the probability that a dietary supplement $p$ contains TiO$_2$.

\subsubsection{Medicines}

The exposure \emph{frequency} of medicines containing TiO$_2$ was modeled using the HIS and FAMPH data (Health Interview Survey data and Federal Agency for Medicines and Health Products data, see Table~\ref{tab1}) as $y^{M_f}_{i} \sim \mbox{Bernoulli}(\pi^{M_f}_i)$, where $y^F_{m_p}$ is a binary indicator of whether individual $i$ is a regular consumer of medicines containing TiO$_2$ (0: no; 1: yes) and $\pi^{M_f}_i$ is the corresponding probability.\\

The exposure \emph{amount} of medicines containing TiO$_2$ was modeled using the HIS data as follows

\begin{align}
\frac{({y^{M_a}_i / y^{M_w}_i)}^{\lambda^M} - 1} {\lambda^M} &\sim \mbox{Normal} (\mu^{M_{a/w}}_i, \sigma^{M_{a/w}}) \hspace{1cm}  \mbox{ for } i = 1, \dots, N^{M_a} \label{eq11}\\
\mu^{M_{a/w}}_i &= \phi^0_i + \sum_{k = 1}^K \phi^k_{l_k[i]}   \label{eq12}
\end{align}
where $y^{M_a}_i$ is the total daily amount of medicine containing TiO$_2$ consumed by individual $i$ (in mg) given that $y^{M_f}_i = 1$, such as to only include non-zero values;
$y^{M_w}_i$ is the individual body weight, which was modeled similarly to (\ref{eq7}); and $\phi^0$ and $\phi^k_{l_k}$ have a similar interpretation to (\ref{eq5}), for $G = 1$. A Box-Cox transformation of the total daily amount $y^{M_a}_i$ adjusted for body weight $y^{M_w}_i$ was assumed to follow a Normal distribution given $\lambda^M$. Since the HIS data does not contain explicit information on the actual amount of each medicine individuals consumed daily, the following adjustment was incorporated in the model to approximate the total daily amount
\begin{align}
y^{M_a}_i &=  \sum y^{M_w}_{p[i]} \label{eq13}\\
y^{M_w}_{p[i]} &\approx u_p \label{eq14}\\
u_p & \sim \mbox{Uniform} (0.8, 2.6) \label{eq15}
\end{align}
where $y^{M_w}_p$ is the amount of TiO$_2$ in medicine $p$, which was approximated by randomly drawing a value from a uniform distribution.  One dispensing unit of a medicine in solid oral dosage formulations (e.g., tablet) has been shown to contain 1.4 (0.8 -- 2.6) mg of TiO$_2$ \cite{cairat_2024}. The bounds of the distribution were therefore selected to be consistent with these findings. Since individuals may consume more than one medicine per day, the approximated medicine weights were summed per individual to compute the total daily amount consumed. This was repeated in each iteration of the MCMC algorithm to account for the intrinsic uncertainty of it.

In contrast to the food and dietary supplements, this approach to estimating the exposure frequency and amount inherently incorporates information on the \emph{market presence} and \emph{concentration} of TiO$_2$ in medicines, therefore bypassing the need to model them separately.

\subsubsection{Personal care products}

Due to a lack of data on the usage of personal care products in Belgium, the exposure \emph{frequency} and \emph{amount} of toothpaste and lip balm could not be modeled at the individual level. Since a large-scale survey on the subject was performed in the neighboring country France, consumption in the French population was assumed to be representative of that in the Belgian one. Although the resulting PCP data (Personal care product survey data, see Table~\ref{tab1}) is not publicly available, the related report includes point estimates of the proportion of consumers and the median daily amount consumed. These were used as constant terms during post-processing, with $\mu^{P_a}_{gi}$ and $\pi^{P_f}_{gi}$ respectively representing the exposure amount --- adjusted for body weight --- and exposure frequency for individual $i$ in category $g$ varying by demographic group. Following the recommendations of the Scientific Committee on Consumer Safety \cite{sccs_2023}, a retention factor $r$ (i.e. the fraction of the product ingested after exposure) of 1.00 was assumed for lip balm in all age groups and 0.05 for toothpaste in adults. Since children tend to ingest between 11\% and 67\% of toothpaste when brushing their teeth \cite{petrovic_2023}, a variable retention factor of toothpaste $e_i \sim$ Uniform(0.11, 0.67) was assumed between the ages of 3 and 9.\\

Similarly to dietary supplements and medicines, the \emph{concentration} of TiO$_2$ in personal care products was not explicitly modeled.\\

The \emph{market presence} of TiO$_2$ in personal care products was modeled using the FAR and the SUP data (Farmaline and supermarket data, see Table~\ref{tab1}), similarly to the one described for the food, but adapted to the case of $G=2$, i.e. toothpaste or lip balm.

\subsubsection{Aggregated exposure assessment} \label{sec3.2.5}

Following modeling, a simulation-based approach was used to generate pseudo-populations reflecting the full spectrum of individual exposure scenarios to TiO$_2$ from food, dietary supplements, medicines, and personal care products. A post-stratification scheme was implemented to ensure representativeness of the Belgian population regarding the long-term aggregated TiO$_2$ exposure for the observed variables \emph{age}, \emph{gender}, and \emph{province}. In each iteration of the MCMC algorithm, a number of pseudo-individuals were sampled per demographic group proportional to the corresponding number in the population according to the DEM data (Statbel data, see Table~\ref{tab1}), with a total of 100,000. For each one, a value of the long-term aggregated exposure was determined by randomly sampling and computing values for the exposure amount, frequency, concentration, and market presence per exposure source. While the steps described at the beginning of Section~\ref{sec2} serve as a general guideline, a more detailed description of the algorithm tailored to the case of TiO$_2$ is given in Appendix B.

\subsection{Software implementation} \label{sub:stan}

All statistical analyses were performed using R version 4.4.1 \cite{r}.
The modeling in Stan was implemented with the CmdStanR package \cite{cmdstanr}, and the generating of pseudo-populations a posteriori was facilitated by the parallel package, which supports parallelized processing by leveraging multiple processors or cores, effectively reducing computation time. To achieve a better performance, a 128-core GPU made available by the Data Science Research Infrastructure hosted at Maastricht University was used.

The corresponding code and datasets are available on GitHub at \url{https://github.com/sophievdn/Aggregated-exposure-assessment-TiO2}.

\subsection{Results}

\begin{figure}[H]
    \centering
    \includegraphics[scale=0.36]{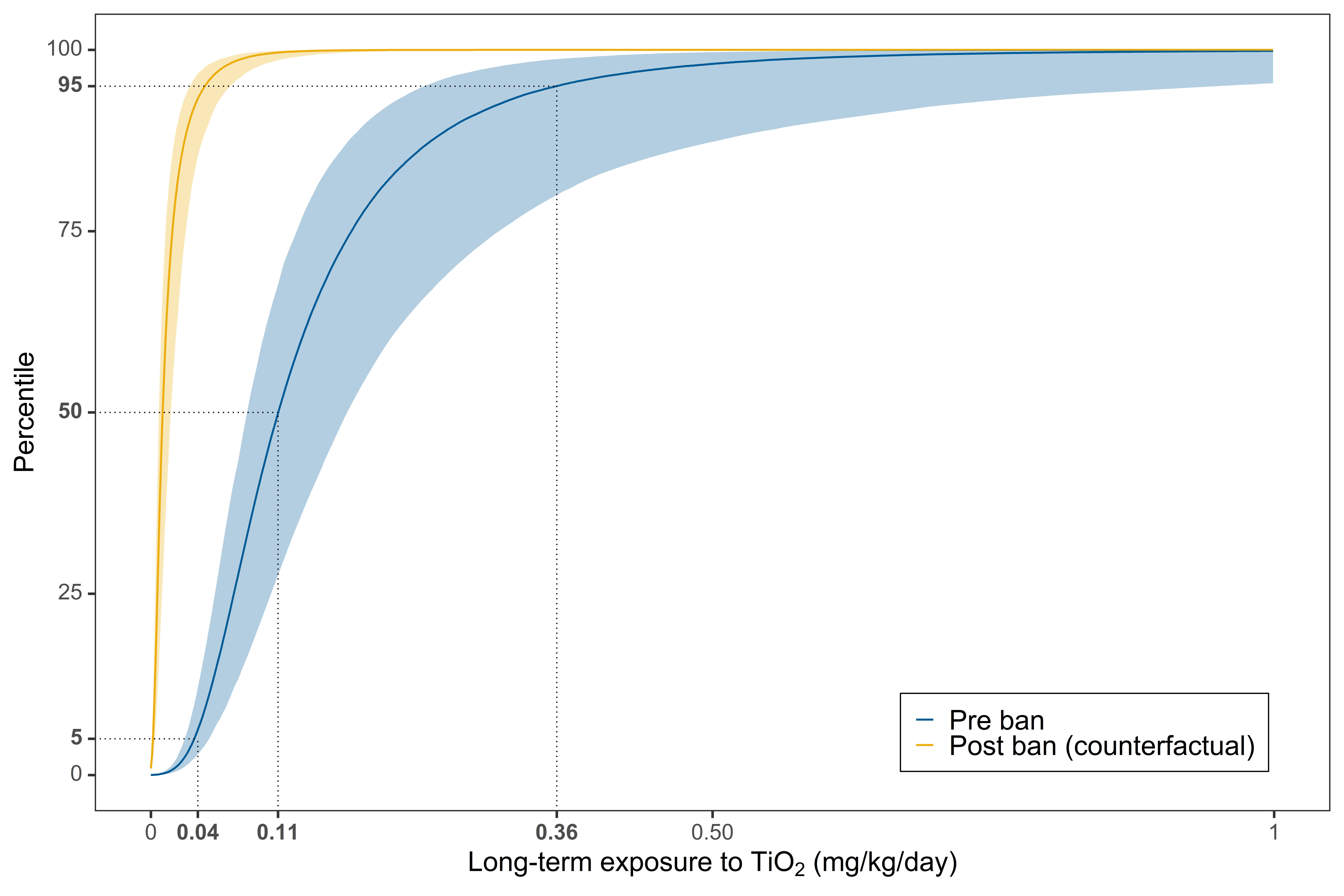}
    \caption{The estimated cumulative distribution function of the long-term aggregated exposure to TiO$_2$ from ingestion in the Belgian population before and after its 2022 ban as a food additive in the EU --- expressed in mg of TiO$_2$ ingested per kg of body weight per day. The solid line represents the posterior median and the shaded area the corresponding 95\% credible region.}
    \label{fig2}
\end{figure}

By generating pseudo-populations in a Bayesian simulation-based approach, a distribution was obtained quantifying both uncertainty and variability in the long-term aggregated exposure to TiO$_2$ via ingestion in the Belgian population --- shown in Figure~\ref{fig2}. It is expressed as the average daily amount of TiO$_2$ that individuals are exposed to relative to their body weight and represents the aggregation of the four exposure sources that were considered, namely food, dietary supplements, medicines, and personal care products. 

A distinction was made between the exposure before and after the 2022 ban of TiO$_2$ as a food additive in the EU in order to assess the impact of this regulatory change. Before the ban, the median long-term exposure was estimated at 0.1139 (0.0881 -- 0.1732) mg/kg/day, representing a typical level of exposure, and the 95th percentile at 0.3624 (0.2445 -- 0.9382) mg/kg/day, representing the level not exceeded by most individuals in the population. The latter is in principle used to determine whether the exposure is within a certain limit that is considered acceptable without being harmful to human health. However, no such limit has been determined for titanium dioxide via ingestion due to the inconclusiveness of the scientific evidence regarding its toxicity \cite{efsa_2016}. Previous exposure assessments of TiO$_2$ have reported average levels ranging from 0.06 to 10.50 mg/kg/day across demographic groups \cite{weir_2012, sprong_2015, rompelberg_2016, efsa_2016}. However, these were based on the assumption that 100\% of products in a particular category contain TiO$_2$, therefore ignoring any information on market presence. This is notwithstanding the fact that these assessments were performed using different methodological approaches and focusing on other countries and time periods, making direct comparison questionable. After the ban, the median and 95th percentile of the resulting distribution were estimated at 0.0108 (0.0075 -- 0.0178) mg/kg/day and 0.0483 (0.0350 -- 0.0707) mg/kg/day, respectively. Since the ban concerned food and dietary supplements, these estimates were derived from the aggregated exposure encompassing only medicines and personal care products ---  specifically, toothpaste and lip balm. Assuming that the consumption and usage patterns did not vary over time, the exposure to these sources before the ban was used as a counterfactual proxy for it after the ban. The significant decrease in exposure is explained by the fact that food was no longer considered in the estimation of the exposure after the ban, though it was the main contributing source to the aggregated exposure before it, representing over 87 (81 -- 93) \% by weight. The main contributing source after the ban was actually found to be medicines, representing over 81 (75 -- 86) \%  by weight across demographic groups for a median of 0.0090 (0.0086 -- 0.0095) mg/kg/day. This is in accordance with a recent exposure assessment of TiO$_2$ \cite{cairat_2024}. Nevertheless, the actual contribution of each source at the individual level depends on an individual's consumption and usage habits. As such, it is possible that the ban had no impact on the extent to which some are exposed to TiO$_2$.\\

\begin{figure}[H]
    \centering
    \includegraphics[scale=0.36]{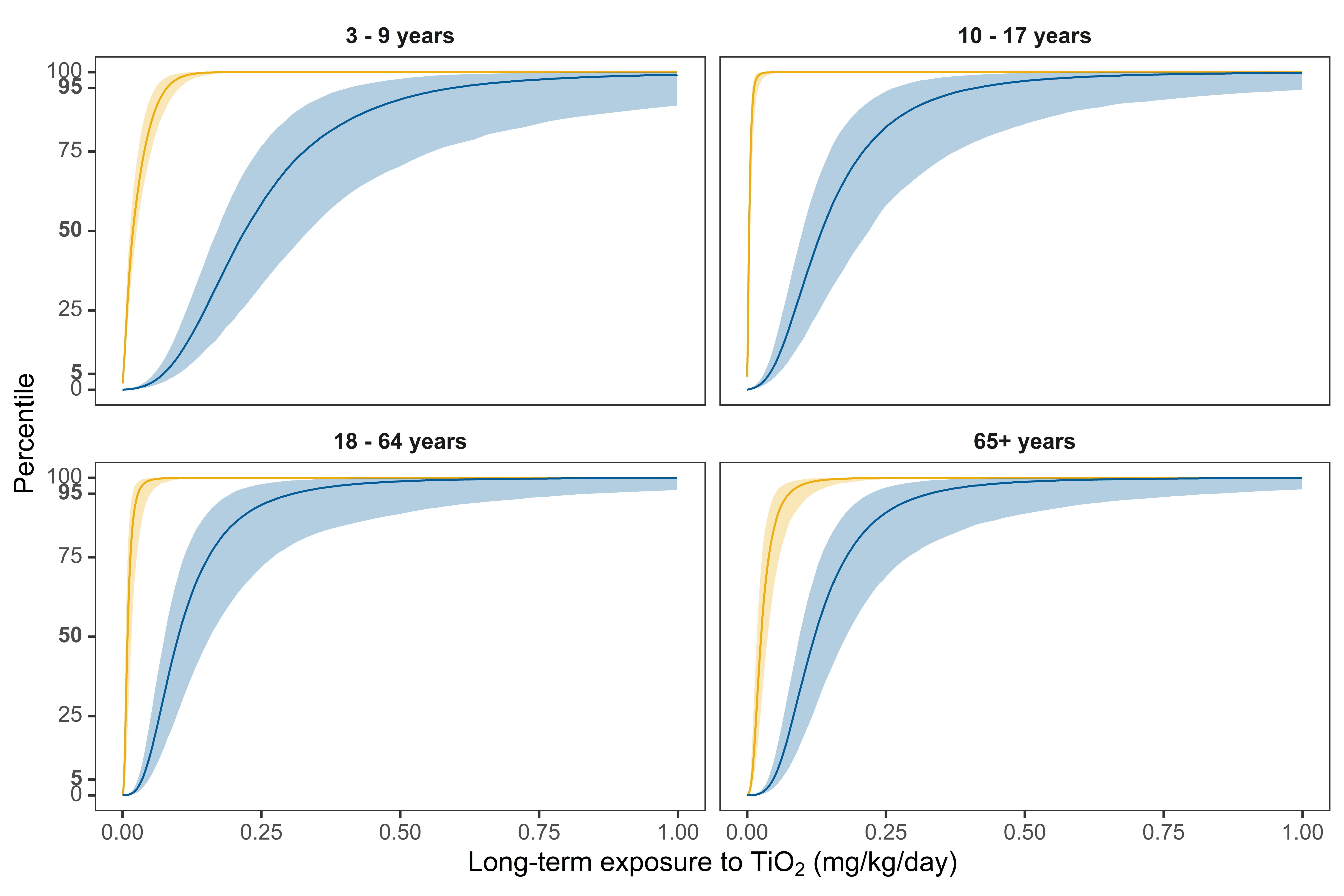}
    \caption{The estimated cumulative distribution function of the long-term aggregated exposure to TiO$_2$ from ingestion in the Belgian population before (\emph{blue}) and after (\emph{yellow}) its 2022 ban as a food additive in the EU by age group --- expressed in mg of TiO$_2$ ingested per kg of body weight per day. The solid line represents the posterior median and the shaded area the corresponding 95\% credible region.}
    \label{fig3}
\end{figure}

In order to identify which demographic groups in the Belgian population are most vulnerable to exposure to TiO$_2$ from ingestion, the median long-term exposure was also estimated by age and gender for and across each source --- given in Table~\ref{tab3}. Children under the age of 10 were found to be the most exposed, irrespective of gender, before the ban --- at the aggregated level and also at the source-specific level for food and personal care products. Conversely, individuals in higher age groups typically had a lower exposure, as shown in Figure~\ref{fig3}. Incidentally, children consume more confectionery and bakery wares, which are the food categories found to contain the most products with TiO$_2$. 
These findings are in accordance with previous exposure assessments of TiO$_2$ performed in different settings and using different methodological approaches \cite{weir_2012,sprong_2015, rompelberg_2016}. Indeed, children are generally more susceptible to chemical exposure than adults due to their smaller physical size and different consumption or usage habits.  However, an opposite cross-sectional trend was found after the ban at the aggregated level, with adults over the age of 64 being the most vulnerable demographic group, irrespective of gender, due to their higher consumption of medicines. This is also in accordance with previous findings \cite{cairat_2024}. \\

When considering that the type of TiO$_2$ used in food, dietary supplements, and medicines in the EU, namely `E 171', has been found to contain a mass percentage of nanoparticles ranging from 3\% to 41\% \cite{verleysen_2022}, it is of interest to determine the extent of exposure in the population to TiO$_2$ in its nanoparticulate form. Including this information probabilistically at the pseudo-individual level yielded a distribution with a median of 0.0274 (0.0210 -- 0.0395) mg/kg/day and a 95th percentile of 0.0819 (0.0611 -- 0.1670) mg/kg/day before the ban. This is considerably higher than previously reported estimates below 0.0001 mg/kg/day, but is explained by the fact that those were obtained assuming a lower mass percentage of nanoparticles, namely 0.31\% \cite{rompelberg_2016}.\\

\begin{sidewaystable} 
\caption{The estimated long-term exposure to TiO$_2$ from ingestion in the Belgian population by age group and gender per exposure source --- expressed in mg of TiO$_2$ ingested per kg of body weight per day. The posterior median of the parameter of interest is reported with the upper and lower bounds of the corresponding 95\% credible interval.}
\label{tab3}
\begin{tabular*}{\textwidth}{@{\extracolsep{\fill}}lcccccc@{\extracolsep{\fill}}}

    \toprule
    
         & \textbf{Food} & \textbf{Dietary supplements} & \textbf{Medicines} & \textbf{Personal care products} \\

    \midrule
\textbf{3-9 years} &&&&\\

    \hspace{0.5 cm}   \textbf{Female}  & 0.1788 (0.1293 -- 0.2816) & 0.00001 (0.00001 -- 0.00006)  & 0.0030 (0.0018 -- 0.0057) & 0.0092 (0.0061 -- 0.0146) \\

    \hspace{0.5 cm}   \textbf{Male}  & 0.1967 (0.1426 -- 0.3176) & 0.00003 (0.00001 -- 0.00015) & 0.0023 (0.0014 -- 0.0043.) & 0.0110 (0.0073 -- 0.0167) \\

    &&&&  \\

    \textbf{10-17 years} &&&&\\
    
    \hspace{0.5 cm}   \textbf{Female}  & 0.1178 (0.0849 -- 0.1917) & 0.00001 (0.00001 -- 0.00006) & 0.0025 (0.0016 -- 0.0048) & 0.0013 (0.0009 -- 0.0019) \\
    
    \hspace{0.5 cm}   \textbf{Male}  & 0.1299 (0.0943 -- 0.2107) & 0.00004 (0.00001 -- 0.00016) & 0.0018 (0.0013 -- 0.0035) & 0.0008 (0.0006 -- 0.00119) \\

    &&&&  \\

    \textbf{18-64 years} &&&&\\
    
    \hspace{0.5 cm}   \textbf{Female}  & 0.0793 (0.0573 -- 0.1312) & 0.00001 (0.00001 -- 0.00006) & 0.0092 (0.0057 -- 0.0168) & 0.0011 (0.0009 -- 0.0014) \\
    
    \hspace{0.5 cm}   \textbf{Male}  & 0.0886 (0.0639 -- 0.1448) & 0.00004 (0.00001 -- 0.00015) & 0.0067 (0.0041 -- 0.0131) & 0.0007 (0.0005 -- 0.0009) \\

     &&&&  \\

    \textbf{65+ years} &&&&\\

    \hspace{0.5 cm}   \textbf{Female}  & 0.0799 (0.0570 -- 0.1312) & 0.00010 (0.00003 -- 0.00035) & 0.0266 (0.0182 -- 0.0404) & 0.0011 (0.0009 -- 0.0014) \\
    
    \hspace{0.5 cm}   \textbf{Male}  & 0.0892 (0.0638 -- 0.1485) & 0.00004 (0.00001 -- 0.00016) & 0.0204 (0.0140 -- 0.0312) & 0.0007 (0.0005 -- 0.0010)\\

    &&&&  \\
\midrule
    \textbf{Overall} & 0.0922 (0.0666 -- 0.1523) & 0.00003 (0.00001 -- 0.00010) & 0.0082 (0.0052 -- 0.0153) & 0.0010 (0.0008 -- 0.0013)\\
    \bottomrule
\end{tabular*}

\end{sidewaystable}

The previously estimated exposure distributions inherently include information on the presence of TiO$_2$ on the Belgian market. This varies not only between, but also within exposure sources when considering different subcategories. In food, the proportion of products containing TiO$_2$ as an ingredient was estimated to range from 0.18 (0.06 -- 0.41) \% for processed fruit and vegetables to 6.75 (6.10 -- 7.44) \% for confectionery. The market presence was estimated to be larger in dietary supplements with an estimated 21 (17 -- 26) \% and even more so in personal care products, with an estimated 27 (32 -- 37) \% for lip balm and 25 (18 -- 33) \% for toothpaste.  Nevertheless, the largest market presence was found in medicines, with an estimated 67 (66 -- 68) \% of solid oral dosage forms containing TiO$_2$ as an ingredient. Not accounting for this information results in an estimated median aggregated exposure of 1.7820 (1.2174 -- 2.9850) mg/kg/day, which is an increase by a factor of approximately 18. This emphasizes the importance of incorporating market presence information when assessing the exposure of a chemical such as TiO$_2$.

\section{Discussion}

When determining the potential risk that a particular chemical poses to public health, it is essential to understand to what extent individuals in the population are exposed to it from all sources. However, accurately capturing the complexity of real-world exposure patterns requires a highly flexible statistical approach. To this purpose, the present work has demonstrated how the use of Bayesian inference for long-term aggregated exposure assessment can fill in some of the limitations characterizing existing methods in this regard.  By applying this fully probabilistic approach on the chemical TiO$_2$, several of its strengths were highlighted, such as the imputation of missing values (e.g., body weight), the incorporation of external knowledge (e.g., retention factor), the flexible transformation of variables, the estimation of the dependency between factors, the assimilation of different sources of uncertainty, and the leveraging of information across unbalanced groups. Nevertheless, some of its limitations were also identified in the process.\\

The approximated joint posterior distribution that we derive when using Bayesian inference represents our state of knowledge about the underlying data-generating process. However, this knowledge primarily depends on the available data and the corresponding model specification. While this holds true for any statistical analysis, it is particularly pertinent when considering the complexity of aggregated exposure --- manifested by the number of exposure sources and factors that need to be considered, as well as the interplay between them. In the absence of one data source encompassing all of these at the individual level, distinct data sources need to be retrieved --- nine in the case of TiO$_2$ --- and linked at the lowest possible hierarchical level (e.g., demographic group, product category). Notwithstanding the fact that qualitative data is often not readily available due to restrictions on access, usage, and distribution. Alternatively, using data of a lesser quality can introduce bias and thus considerably alter any conclusions that are drawn regarding the risk that a chemical poses to human health. For example, when estimating the exposure to TiO$_2$ from medicine, analytical measurements collected from a systematic review were initially used as input to model the related concentration. This approach was eventually disregarded in favor of one based on findings from a study using more representative, albeit not publicly available, data. Shifting from one approach to the other remarkably led to a tenfold increase in the magnitude of exposure.  Furthermore, specifying a model that fits the data well requires a tailored approach based on a profound understanding of statistical analysis and programming --- effectively bringing into question the propriety of so-called `black box' approaches commonly used in public health. \\

Since real data on aggregated exposure in the population often does not exist, a pseudo-population is generated by randomly sampling values from these distinct models and subsequently integrating them. Through an iterative simulation-based approach, the distribution of long-term aggregated exposure is estimated, incorporating both variability and uncertainty. However, this approach relies on a range of plausible assumptions that are defined to fill in the gaps in our knowledge about the underlying data-generating process. Failing to accurately capture reality could make the results unreliable, which is particularly problematic in the context of regulatory decision-making. For example, since we did not have data on the relationship between the different sources of TiO$_2$, independence was assumed. Yet, one could hypothesize that individuals who, say, restrict their diet to certain food categories also consume more dietary supplements. Furthermore, when generating a pseudo-population, a post-stratification approach can be introduced to adjust for non-representativeness in the data based on observed variables (e.g., age). However, this does not account for any bias that might be introduced due to unobserved ones. For example, in a survey about health-related behavior, it is possible that individuals who are less health-conscious are also less likely to respond, leading to an overestimation of how well the population is doing in that regard.\\

From modeling to generating desired quantities in a simulation-based approach, the estimation of aggregated exposure using Bayesian inference has a considerable computational cost that generally increases with complexity. In the case of TiO$_2$, the entire computation lasted approximately six days despite using a high-performance system for part of it.  This is explained by the fact that an MCMC algorithm generally requires a certain amount of time to effectively explore the joint posterior distribution when the number of parameters is large --- over 70,000 for the aggregated exposure of TiO$_2$. However, there are also alternative approaches, such as variational inference and Laplace approximation, which are less computationally expensive in high-dimensional spaces but trade off accuracy for efficiency \cite{blei_2017, tierney_1986}. Nevertheless, the ongoing development of more efficient algorithms and faster computer processors will undoubtedly reduce the computational burden of using Bayesian inference in the future.\\ 

Lastly, the proposed approach assumes exposure assessment to a defined product with a specific chemical composition on a mass base. For chemicals such as TiO$_2$, inherent variation in their physicochemical properties, such as their size and shape distributions, crystallographic phase and purity, can result in different forms of it with different hazards. The relative concentrations of forms and associated exposure levels can vary depending on their application. This is not accounted for in the proposed approach and in conventional risk assessment (\citealp[e.g.,][]{efsa_2021_nano}).\\

Despite the aforementioned limitations, the benefits of using Bayesian inference for the estimation of aggregated exposure far outweigh them. By simultaneously considering multiple sources of exposure, the proposed framework can support more informed regulatory decision-making protecting the public's health, thus moving away from the less realistic traditional `one exposure -- one health effect' approach.  Future research priorities should encompass: (i) incorporation of physicochemical variability; (ii) extension to multichemical exposure scenarios to address real-world mixture exposures; and (iii) integration of advanced computational approaches, including large language models, to enhance data synthesis capabilities and address current methodological limitations. Such developments would further strengthen the scientific foundation for evidence-based environmental health policy and advance toward a more comprehensive exposome characterization.

\pagebreak
\pagebreak
\Urlmuskip=0mu plus 1mu\relax

\bibliography{bibliography.bib}

\pagebreak

\renewcommand \thepart{}
\renewcommand \partname{}
\appendix
\pagenumbering{arabic}
\setcounter{page}{1}
\renewcommand*{\thepage}{A\arabic{page}}
\addcontentsline{toc}{section}{Appendix} 
\part{Appendix} 
\parttoc 
\thispagestyle{empty}

\input{appendix.tex}

\end{document}

%% file: appendix.tex
\pagenumbering{arabic}

\section{Extended mathematical model formulations}

\noindent \textit{\textbf{Food}} 
\begin{alignat*}{2}
y^{F_f}_{git} &\sim \mbox{Bernoulli}(\pi^{F_f}_{gi})  &{} \mbox{ for }{}& i = 1, \dots, N^{F_f} \mbox{; } t \in \{1, 2\} \\
\mbox{logit}(\pi^{F_f}_{gi}) &= \eta^0_g + \sum_{k = 1}^{K^{F_f}} \eta^k_{gl_k[i]}  + \upsilon_{gi} &{} \\
\notag \\[2ex]
\frac{({y^{F_a}_{git} / y^{F_a}_{w_i})}^{\lambda^F} - 1} {\lambda^F} &\sim \mbox{Normal} (\mu^{F_{a/w}}_{gi}, \sigma^{F_{a/w}}) &{}  \mbox{ for }{}& i = 1, \dots, N^{F_a} \mbox{; } t \in \{1, 2\} \\
\mu^{F_{a/w}}_{gi} &= \gamma^0_g + \sum_{k = 1}^{K^{F_a}} \gamma^k_{gl_k[i]} + \nu_{gi} \\
\notag \\[2ex]
\mbox{log}(y^{F_w}_{i}) &\sim \mbox{Normal} (\mu^{F_w}_{i}, \sigma^{F_w}) \\
\mu^{F_w}_{i} &= \beta_0 + \sum_{k = 1}^{K^{F_w}} \beta^{k.(k+1)}_{l_k[i], l_{k+1}[i]}
\notag \\[2ex]
\eta^0_g &\sim \mbox{Normal}(\mu^0_{\eta}, \sigma^0_{\eta}) &{} \mbox{ for }{}& g = 1, \dots, G^{F_{a,f}} \\
\gamma^0_g &\sim \mbox{Normal}(\mu^0_{\gamma}, \sigma^0_{\gamma})\\
\mu^0_{\eta}, \mu^0_{\gamma} &\sim \mbox{Normal}(0, 2.5)\\
\sigma^0_{\eta}, \sigma^0_{\gamma} &\sim \mbox{Normal}^+(0, 1)\\
\eta^k_{gl_k} &\sim \mbox{Normal}(0, \sigma^k_{\eta_g}) &{} \mbox{ for }{}&  k = 1, \ldots, K^{F_f} \\
\gamma^k_{gl_k} &\sim \mbox{Normal}(0, \sigma^k_{\gamma_g}) &{} \mbox{ for }{}&  k = 1, \ldots, K^{F_a}\\
\sigma^k_{\eta_g}, \sigma^k_{\gamma_g}, &\sim \mbox{Normal}^+(0, 1) \\
\upsilon_{gi}, \nu_{gi} &\sim \mbox{Normal}(0, \Sigma_{\upsilon, \nu})  \\
\beta_0^{F_w} &\sim \mbox{Normal}(0, 2.5) \\
\beta^{k.(k+1)} &\sim \mbox{Normal}(0, 1) &{} \mbox{ for }{}& k = 1, \ldots, K^{F_w} \\
\sigma^{F_{a/w}}, \sigma^{F_w} &\sim \mbox{Normal}^+(0, 1) 
\end{alignat*}

\newpage

\begin{align*}
y^{F_m}_{p} &\sim \mbox{Bernoulli}(\pi^{F_m}_{p}) &{}  \mbox{ for }{}& p = 1, \dots, N^{F_m} \\
\mbox{logit}(\pi^{F_m}_{p}) &=\delta_0 + \delta_{g[p]} \\
\delta_0 &\sim \mbox{Normal}(0,2.5)\\
\delta_{g} &\sim \mbox{Normal}(0, \sigma_{\delta}) &{} \mbox{ for }{}& g = 1, \dots, G^{F_m} \\
\sigma_{\delta} &\sim \mbox{Normal}^+ (0, 1) 
\end{align*}
\\
\noindent \textit{\textbf{Dietary supplements}}

\begin{align*}
y^{S_f}_{it} &\sim \mbox{Bernoulli}(\pi^{S_f}_{i})   &{}\mbox{ for }{}& i = 1, \dots, N^{S_f} \mbox{; } t \in \{1, 2\} \\
\mbox{logit}(\pi^{S_f}_{i}) &=\alpha^0 + \sum_{k = 1}^{K^{S_f}} \alpha^k_{l_k[i]} + \tau_i \\
\alpha^0 &\sim \mbox{Normal} (0, 2.5) \\
\alpha^k_{l_k} &\sim \mbox{Normal}(0, \sigma^k_{\alpha}) &{}\mbox{ for }{}& k = 1, \dots, K^{S_f} \\
\tau_i &\sim \mbox{Normal}(0, \sigma_{\tau}) \\
\sigma^k_{\alpha}, \sigma_{\tau} &\sim \mbox{Normal}^+(0, 1) \\
\notag \\[2ex]
\mbox{log}\left (\frac{ \sum_t  y^{S_a}_{it} / T_i}{y^{S_w}_{i}} \right ) &\sim \mbox{Normal} (\mu^{S_{a/w}}_{z[i]}, \sigma^{S_{a/w}}_{z[i]}) &{} \mbox{ for }{}& i = 1, \dots, N^{S_a} \mbox{; } t \in \{1, 2\} \\
\mu^{S_{a/w}}_{z[i]} &= \rho^0_{z[i]} + \sum_{k = 1}^{K^{S_a}} \rho^k_{l_k[i]} \\
z_i &\sim \mbox{Categorical}(\vartheta_{z[i]}) &\mbox{for }& z = 1,.., 3 \mbox{ and } \sum^3_{z = 1} \vartheta_z = 1 
\notag \\[2ex]
\mbox{log}(y^{S_w}_{i}) &\sim \mbox{Normal} (\mu^{S_w}_{i}, \sigma^{S_w})\\
\mu^{S_w}_{i} &= \zeta_0 + \sum_{k = 1}^{{K^{S_w}}} \zeta^{k.(k+1)}_{l_k[i], l_{k+1}[i]}\\
\end{align*}

\begin{align*}
\rho^0_{z} &\sim \mbox{Normal}(0, 2.5) &\mbox{for }& k = 1, \ldots, K^{S_a} \\
\rho^k_{l_k} &\sim \mbox{Normal}(0, \sigma_{\rho}) &\mbox{for }& k = 1, \ldots, K^{S_a} \\
\mbox{log}(\sigma^{S_{a/w}}_{z}) &\sim \mbox{Normal}(0, 2) \\
\zeta_0 &\sim \mbox{Normal}(0, 2.5)  \\
\zeta^{k.(k+1)} &\sim \mbox{Normal}(0, 1) &{} \mbox{ for }{}& k = 1, \ldots, K^{S_w} \\
\sigma^{S_w}, \sigma_{\rho} &\sim \mbox{Normal}^+(0, 1) 
\end{align*}

\begin{align*}
y^{S_m}_{p} &\sim \mbox{Bernoulli}(\pi^{S_m}_{p}) &{}  \mbox{ for }{}& p = 1, \dots, N^{S_m} \\
\mbox{logit}(\pi^{S_m}_{p}) &\sim \mbox{Normal} (0, 2.5) 
\end{align*}
\\
\noindent \textit{\textbf{Medicines}}

\begin{align*}
\frac{({y^{M_a}_i / y^{M_w}_i)}^{\lambda^M} - 1} {\lambda^M} &\sim \mbox{Normal} (\mu^{M_{a/w}}_i, \sigma^{M_{a/w}}) &{}  \mbox{ for } {}& i = 1, \dots, N^{M_a} \\
\mu^{M_{a/w}}_i &= \phi^0_i + \sum_{k = 1}^K \phi^k_{l_k[i]} \\
y^{M_a}_i &=  \sum y^{M_w}_{p[i]} \\
y^{M_w}_{p[i]} &\approx u_p\\
u_p & \sim \mbox{Uniform} (0.8, 2.6) \\
\notag \\[2ex]
\mbox{log}(y^{M_w}_i) &\sim \mbox{Normal} (\mu^{M_w}_i, \sigma^{M_w}) \\
\mu^{M_w}_i &= \omega_0 + \sum_{k = 1}^{K} \omega^{k.(k+1)}_{l_k[i], l_{k+1}[i]} 
\end{align*}

\begin{align*}
\phi^0, \omega^0 &\sim \mbox{Normal}(0, 2.5)\\
\phi^k_{l_k} &\sim \mbox{Normal}(0, \sigma^k_{\phi}) &{}  \mbox{ for } {}& k = 1, \ldots, K^{M_a}\\
\omega^{k.(k+1)}_{l_k, l_{k+1}}  &\sim \mbox{Normal}(0, 1) &{} \mbox{ for } {}& k = 1, \ldots, K^{M_w}\\
\sigma^{M_{a/w}}, \sigma^{M_w}, \sigma^k_{\phi} &\sim \mbox{Normal}^+(0, 1)
\end{align*}
\\
\noindent \textit{\textbf{Personal care products}}

\begin{align*}
y^{P_m}_{p} &\sim \mbox{Bernoulli}(\pi^{P_m}_{p}) &{} \hspace{0.5cm} \mbox{ for }{}& p = 1, \dots, N^{P_m} \\
\mbox{logit}(\pi^{P_m}_{p}) &=\xi_0 + \xi_{g[p]} \\
\xi_0 &\sim \mbox{Normal}(0,2.5)\\
\xi_{g} &\sim \mbox{Normal}(0, \sigma_{\xi}) &{} \mbox{ for }{}& g = 1, \dots, G^{P_m} \\
\sigma_{\xi} &\sim \mbox{Normal}^+ (0, 1) 
\end{align*}

\newpage
\section{Algorithm to generate pseudo-populations}

\begin{algorithm}[H]
\caption{}\label{alg:sim}
\begin{algorithmic}[1]
\footnotesize
\Require{A dataset consisting of the sampled values for all the parameters $\theta$ at every iteration of the sampling phase of the MCMC algorithm}
\For {iteration $j = 1, \dots, J$} 
    \For {demographic group $d = 1, \dots, D$} 
        \State{Let level $l$ of predictor $k$ correspond to demographic group $d$};
        \State{Let $N_d$ be the size of the population in demographic group $d$};
        \For{pseudo-individual $i = 1, \dots, N_d$}
        
        \Algphase{Food}
        
            \For{food category $g = 1, \dots, G$}
            \State {Draw $\upsilon_{gi}^{(j)}, \nu_{gi}^{(j)} \sim \mbox{Normal}(0, \Sigma^{(j)}_{\upsilon, \nu})$ ($n=1$)}; \Comment{Subject effects}
            \State {Draw $f(y_{git}^{F_{a/w}^{(j)}}) \sim$ Normal($\gamma^{0^{(j)}}_g + \sum_{k = 1}^K \gamma^{k^{(j)}}_{gl_k[i]} + \nu_{gi}^{(j)}, \sigma^{F_{a/w}^{(j)}}$) ($n=100$)}; 
            \State {Back-transform $f(y_{git}^{F_{a/w}^{(j)}})$} according to:
            \Indent
                \If{$\lambda^{F^{(j)}} \neq 0$}
                    \State{$(\lambda^{F^{(j)}} \mbox{ x } f(y_{git}^{F_{a/w}^{(j)}}) + 1)^{(1/\lambda^{F^{(j)}})}$}
                \Else
                    \State{log$^{-1} (f(y_{git}^{F_{a/w}^{(j)}}))$}
                \EndIf
            \EndIndent
            \State {Compute $\Bar{y}_{gi}^{F_{a/w}^{(j)}} = \sum_t y_{git}^{F_{a/w}^{(j)}} / n $}; \Comment{Amount (adjusted)}
            \State{Compute $\Bar{y}_{gi}^{F_f^{(j)}} = \mbox{logit}^{-1}(\eta^{0^{(j)}}_g + \sum_{k = 1}^K \eta^{k^{(j)}}_{gl_k[i]}  + \upsilon_{gi}^{^{(j)}})$}; \Comment{Frequency}

            \State{Set $\Bar{y}_{gi}^{F_c^{(j)}} = \tilde{\mu}_g^{F_c^{(j)}}$}; \Comment{Concentration}
            \State{Draw $\Bar{y}_{gi}^{F_m^{(j)}} \sim \mbox{PERT}(0,\pi^{F_m^{(j)}}_g,1)$ ($n=1$)}; \Comment{Market presence}
            \State{Compute $\Bar{y}_{gi}^{F^{(j)}} = \Bar{y}_{gi}^{F_{a/w}^{(j)}} \mbox{ x } \Bar{y}_{gi}^{F_f^{(j)}} \mbox{ x } \Bar{y}_{gi}^{F_c^{(j)}} \mbox{ x } \Bar{y}_{gi}^{F_m^{(j)}} \mbox{ x } 10^{-3}$} ; \Comment{Combination of factors}
            \EndFor
            \State{Compute $\Bar{y}_{i}^{F^{(j)}} = \sum_g \Bar{y}_{gi}^{F^{(j)}}$};  \Comment{Sum across categories} 
            
        \Algphase{Dietary supplements}
        
        \State{Draw $z_i^{(j)} \sim \mbox{Categorical}(\vartheta^{(j)}_z)$} ($n=1$); \Comment{Mixture component}
        \State{Draw $f(\Bar{y}_i^{S_{a/w}^{(j)}}) \sim \mbox{Normal}(\mu^{S_{a/w}^{(j)}}_{z[i]}, \sigma^{S_{a/w}^{(j)}}_{z[i]})$} ($n=1$); 
        \State{Compute $\Bar{y}_i^{S_{a/w}^{(j)}} = \mbox{log}^{-1}(f(\Bar{y}_i^{S_{a/w}^{(j)}}))$}; \Comment{Amount (adjusted)}
        \State{Draw $\tau_i^{(j)} \sim \mbox{Normal}(0, \sigma^{(j)}_{\tau})$ ($n=1$)}; \Comment{Subject effect}

\algstore{bkbreak}
\end{algorithmic}
\end{algorithm}

\begin{algorithm}[H]
\begin{algorithmic}[1]
\footnotesize
\algrestore{bkbreak}

        \State{Compute $\Bar{y}_i^{S_f^{(j)}} = \mbox{logit}^{-1}(\alpha^{0^{(j)}} + \sum_{k = 1}^K \alpha^{k^{(j)}}_{l_k[i]} + \tau_i^{(j)})$}; \Comment{Frequency}
        \State{Let $\bm{r}$ be a vector of the observed number of dietary supplements consumed}
        \State{\hspace{0.5cm} per person per day in demographic group $d$;}
        \State{Let $\bm{c}$ be a vector of the observed TiO$_2$ concentrations in dietary supplements};
        \State{Draw $r_i^{(j)}$ with replacement from $\bm{r}$ ($n=1$)};
        \State{Draw $c_i^{(j)}$ with replacement from $\bm{c}$ ($n=r_i^{(j)}$)};
        \State{Compute $\Bar{y}_i^{S_c} = \frac{\sum c_i^{(j)}}{r_i^{(j)}}$}; \Comment{Concentration}
        \State{Draw $\Bar{y}_i^{S_m^{(j)}} \sim \mbox{PERT}(0,\pi^{S_m^{(j)}}, 1)$} ($n=1$); \Comment{Market presence}
        \State{Compute $\Bar{y}_i^{S^{(j)}} = \Bar{y}_i^{S_{a/w}^{(j)}} \mbox{ x } \Bar{y}_i^{S_f^{(j)}} \mbox{ x } \Bar{y}_i^{S_c^{(j)}} \mbox{ x } \Bar{y}_i^{S_m^{(j)}} \mbox{ x } 10^{-3}$} \Comment{Combination of factors}
        
        \Algphase{Medicines}
        \State{Draw $f(\Bar{y}_i^{M_{a/w}^{(j)}}) \sim \mbox{Normal} (\mu_i^{M_{a/w}^{(j)}}), \sigma^{M_{a/w}^{(j)}})$ ($n=1$)}  \Comment{Amount (adjusted)};
        \State{Back-transform $f(\Bar{y}_i^{M_{a/w}^{(j)}})$ according to lines 10--14};
        \State{Draw $\Bar{y}_i^{M_f^{(j)}} \sim \mbox{PERT}(0, \pi^{M_f^{(j)}}_i, 1)$} ($n=1$); \Comment{Frequency}
        \State{Compute $\Bar{y}_i^{M^{(j)}} = \Bar{y}_i^{M_{a/w}^{(j)}} \mbox{ x } \Bar{y}_i^{M_f^{(j)}}   \mbox{ x } 10^{-6}$} \Comment{Combination of factors}

        \Algphase{Personal care products}
            \For{personal care category $g = 1, \dots, G$}
            \If{personal care category $g$ is lip balm} \Comment{Retention factor}
                \State{$e_i^{(j)}=1$}
            \Else {
                \If{age in demographic group $d$ $\leq 12$ years}
                    \State{Draw $e_i^{(j)} \sim \mbox{Uniform}(0.26, 0.67)$ $(n=1)$}
                \Else
                    \State{$e_i^{(j)} = 0.05$}
                \EndIf
            }
            \EndIf 
            \State{Compute $ \Bar{y}_{gi}^{P_{a/w}} = \mu^{P_a^{(j)}}_{gi} \mbox{ x } e_i^{(j)}$}; \Comment{Amount (adjusted)}
            \State{Draw $\Bar{y}_{gi}^{P_f} \sim \mbox{PERT}(0, \pi^{P_f^{(j)}}_{gi}, 1)$ $(n=1)$}; \Comment{Frequency}
            \State{Compute $\Bar{y}_i^{P_c^{(j)}}$ according to lines 26-31} \Comment{Concentration}
            \State{Draw $\Bar{y}_{gi}^{P_m^{(j)}} \sim \mbox{PERT}(0,\pi_g^{P_m^{(j)}},1)$ ($n=1$)}; \Comment{Market presence}
            \State{Compute $\Bar{y}_{gi}^{P^{(j)}} = \Bar{y}_{gi}^{P_{a/w}^{(j)}} \mbox{ x } \Bar{y}_{gi}^{P_f^{(j)}} \mbox{ x } \Bar{y}_{gi}^{P_c^{(j)}} \mbox{ x } \Bar{y}_{gi}^{P_m^{(j)}}  \mbox{ x } 10^{-6}$}; \Comment{Combination of factors}
            \EndFor
            \State{Compute $\Bar{y}_i^{P^{(j)}} = \sum_g \Bar{y}_{gi}^{P^{(j)}} $}; \Comment{Sum across categories} 

        \Algphase{Aggregation across exposure sources}
        \State{Compute $\Bar{y}_i^{(j)} = \Bar{y}_i^{F^{(j)}} + \Bar{y}_i^{S^{(j)}} + \Bar{y}_i^{M^{(j)}} + \Bar{y}_i^{P^{(j)}}$}
        \EndFor
    \EndFor
\EndFor
\end{algorithmic}
\end{algorithm}